# Effects of the space plasma density oscillation on the inter-spacecraft laser ranging for TianQin gravitational wave observatory

Ling-Feng Lu[1], Wei Su[2*], Xuefeng Zhang[1], Zhao-Guo He[3], Hui-Zong Duan[1], Yuan-Ze Jiang[2], Hsien-Chi Yeh[1]

[1]TianQin Research Center for Gravitational Physics and School of Physics and Astronomy, Sun Yat-sen University (Zhuhai Campus), Zhuhai 519082, P. R. China, [2]MOE Key Laboratory of Fundamental Physical Quantities Measurement, & Hubei Key Laboratory of Gravitation and Quantum Physics, PGMF and School of Physics, Huazhong University of Science and Technology, Wuhan 430074, P. R. China, [3]School of Atmospheric Sciences, Sun Yat-sen University (Zhuhai Campus), Zhuhai 519082, P. R. China

*Corresponding author: Wei Su (suw12@hust.edu.cn)

**Key points**

- Additional phase deviation of the laser beams induced by the space plasmas is estimated for TianQin gravitational wave observatory
- The oscillating plasma density could cause the phase deviation close to $2\times10^{-6}$ rad/Hz$^{1/2}$ or 0.3pm/Hz$^{1/2}$ in the mHz frequency band
- The plasma influence becomes more moderate when the constellation plane is perpendicular to the Sun-Earth line and with low orbital radius

**Abstract**

The TianQin space Gravitational Waves (GW) observatory will contain 3 geocentric and circularly orbiting spacecraft with an orbital radius of $10^5$ km, to detect the GW in the milli-hertz frequency band. Each spacecraft pair will establish a $1.7\times10^5$ km-long laser interferometer immersed in the solar wind and the magnetospheric plasmas to measure the phase deviations induced by the GW. GW detection requires a high-precision measurement of the laser phase. The cumulative effects of the long distance and the periodic oscillations of the plasma density may induce an additional phase noise. This paper aims to model the plasma induced phase deviation of the inter-spacecraft laser signals, using a realistic orbit simulator and the Space Weather Modeling Framework (SWMF) model. Preliminary results show that the plasma density oscillation can induce the phase deviations close to $2\times10^{-6}$ rad/Hz$^{1/2}$ or 0.3pm/Hz$^{1/2}$ in the milli-hertz frequency band and it is within the error budget assigned to the displacement noise of the interferometry. The amplitude spectrum density of phases along three arms become more separated when the orbital plane is parallel to the Sun-Earth line or during a magnetic storm. Finally, the dependence of the phase deviations on the orbital radius is examined.

## 1 Introduction

In 2015, the LIGO team carried out the first direct detection of Gravitational Waves (GW) (Abbott et al., 2016). The ground-based GW detectors are sensitive to the frequency band from 10Hz to 1000Hz. In contrast, the targeted frequency band of the space-borne GW detection programs is around milli-hertz. The principle of the GW detection is straightforward. Due to the polarizations of the GW, when a GW passes the GW observatory, one expects to see one arm expands, while the other shrinks. Just like the ground-based GW detectors (Abbott et al., 2016), the space borne GW detectors make use of the laser interferometers to measure the tiny length variations in-between the test masses inside each spacecraft. Laser beams are emitted and received through two telescopes installed in each spacecraft. The variation of the path length of the laser beam is converted to the phase variation and then being measured. LISA (Vitale, 2014), Taiji (Hu, W. R. & Wu, 2017), DECIGO (Kawamura et al., 2011) and TianQin (Luo et al., 2016) are among the well-known ongoing space-borne GW observatories. The first two adopt the heliocentric orbit, whereas the latter two follow the geocentric orbit. TianQin is a Chinese space borne gravitational wave detection mission (Luo et al., 2016). It is envisaged to launch 3 circularly orbiting spacecraft at a distance of $10^5$ km (~15.6 $R_E$, $R_E$ is the Earth radius) from the center of the Earth in the early 2030s. The three spacecraft will constitute an equilateral triangular constellation, with each arm having a length of $1.7\times10^5$ km. Currently, the normal of the orbit plane is designed to point at a well-chosen GW source, the white-dwarf binary RX J0806.3+1527 (hereafter J0806) which emits

GWs in a period of 2.7 mins (or 6 mHz), although other milli-hertz sources are also being considered. Figure 1 shows the orbit of the TianQin constellation (Lu et al., 2020) and its position with respect to the ecliptic plane. Following the rotation of the Earth, four three-months-long periods can be distinguished, AB, BC, CD and DA, respectively. Under the current operation scheme, the GW detection will be mainly conducted during the periods of AB and CD due to the fact that during the period of AD and BC, the telescopes installed in the spacecraft will face sunlight directly, which is detrimental to the laser ranging experiment.

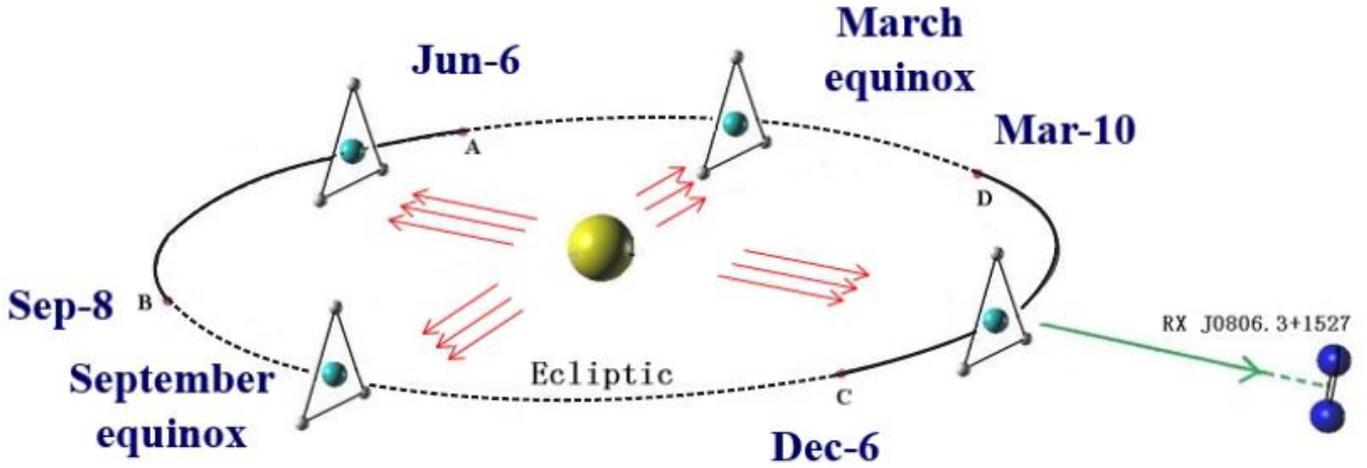

Figure 1. The orbit of the TianQin constellation and the ecliptic plane. A-D specified the approximating dates separating each operation period. The dashed lines indicate the non-detection periods, whereas the solid lines depict the detection periods. The green arrow points to the location of the J0806 GW source. Figure cited from Lu et al. (2020).

GW detection in space adopts extremely high precision laser interferometers. The overall accuracy requirement of the displacement measurement is roughly proportional to the arm length. For TianQin, provided the arm length is $1.7\times10^5$ km, the displacement measurement accuracy $S_{dis}$ should fulfill the following empirical criterion (Luo et al., 2016).

$$S_{dis}^{1/2} \leq \frac{10^{-12}}{\sqrt{2}} \frac{m}{\sqrt{Hz}} \sqrt{1+(\frac{10mHz}{f})^4} \qquad (1)$$

where $f$ is the frequency, the numerator 10 mHz is determined by the interplay of the acceleration accuracy, the arm length and the position accuracy. Essentially, Eq. (1) means the precision of the displacement measurement should reach 1pm/Hz$^{1/2}$ at 10mHz. Unlike the ground based GW detector, the space GW detection is conducted through the space plasmas. The error of the laser path attributed to space plasmas should be no more than 30% of the overall accuracy requirement. Accordingly, the phase shift of the laser beams due to plasma effects after propagating $1.7\times10^5$ km should be less than $\frac{2\pi \times 30\% \times 10^{-12}}{\sqrt{2}\lambda} \frac{rad}{\sqrt{Hz}} \sqrt{1+(\frac{10mHz}{f})^4}$, where $\lambda$=1064 nm is the wavelength of the adopted Nd:YAG laser.

Following the orbit of the spacecraft, the laser beam in between two spacecraft will pass through the bow shock, the magnetosheath, the magnetopause and the inner magnetosphere. The major plasma species in these region/boundary are e$^-$, H$^+$, O$^+$, He$^{2+}$, etc. The typical plasma density in these regions are about 1-10 cm$^{-3}$. The particle velocity is about 400 km/s and the magnetic field is 10-100 nT (Baumjohann & Treumann, 1997). In such a weak magnetic field, phase deviations of the laser beam ($\Delta\phi$) traveling through space plasmas can be estimated by the following equation (Hutchinson, 2002),

$$\Delta\phi = \frac{2\pi}{\lambda} \int_0^L \frac{n_e q^2}{2\omega_0^2 \varepsilon_0 m_e} dl \qquad (2)$$

where $L=1.7\times10^5$ km is the arm length, $q$ the unit electric charge, $\omega_0$ the angular frequency of laser, $\varepsilon_0$ the vacuum permittivity, $m_e$ the electron mass and $n_e$ the plasma electron density.

So the question is, given the required high precision, whether the "thin" space plasma could introduce an excessive noise to the phase measurement? More importantly, the GW measurement relies on the detection of the periodically increasing or decreasing of the arm length. There are plenty of physical mechanisms which can induce plasma density oscillations in the same frequency band as the GW frequencies. For example, Viall et al. (2009a, 2009b) analyzed 11 years' particle density and magnetic field data from *in-situ* measurement by WIND (Ogilvie et al., 1995) and GOES (Singer et al., 1996) satellites. They found a couple of discrete frequencies within 0.5-5 mHz occurring more often than others both in the solar wind and the magnetosphere. The origin of these periodic oscillations in the magnetosphere and the solar wind is probably due to the solar wind-magnetopause interaction (Song et al., 1988) or the modulations by the Ultra-Low Frequency (ULF) waves (Samson, 1991). Taking these facts into consideration, one may ask whether the phase deviations given by Eq. (2) also have the periodic oscillations in the milli-hertz frequency band. Moreover, one of the worst space weather the GW detector will experience is the magnetic storm (Vennerstroem, 2001). The magnetic storm could cause a significant increase of the plasma density and thus affect the phase deviation. Although the magnetic storm is, more or less, detectable and one could discard certain data series or put the spacecraft on hold, it is important to know if the interferometry could work under the worst circumstances. Another intriguing question is the difference of the phase deviations in the detection period *i.e.* period AB in Figure 1 and non-detection period, *i.e.* period BC in Figure 1, as the alignment of the orbit affects the relative positions of the spacecraft in the magnetosphere. Finally, how does the phase deviation evolve when changing the orbital radius of the constellation?

Prior to this work, the wavefront distortion of the inter-spacecraft laser beams due to the turbulence of the space plasmas has been analyzed for TianQin (Lu et al., 2020). That work did not consider the phase deviation induced by the line integral of the plasma density. To the knowledge of the authors, there is no other peer-reviewed literatures discussing specifically the plasma induced phase deviations on any space-borne GW detection programs. Through this work, it is hoped that readers would get a more comprehensive impression on whether the plasma density oscillations could affect the GW detection.

This paper aims to investigate quantitatively the effect of the plasma density oscillations on the phase deviation of the laser signals. From Eq.(2), we know that the phase deviation depends on the time-varying density distribution along the laser's propagation path. Since the *in-situ* measurement just give density distributions at discrete points, but not the density distribution along the propagation path, a sophisticated space plasma density model is used, combing the realistic spacecraft orbits of the TianQin constellation, to investigate the above questions. The structure of this paper is the following: section 2 introduces the orbital simulator and the plasma density model. Section 3 simulates the phase deviation of laser beams due to plasma density oscillations at the orbital radius of $10^5$ km. The phase deviations in time domain as well as its power spectra are shown. For the sake of completeness of the simulation, the phase deviations in both the detection period and the non-detection period are studied. Section 4 examines the effects of the orbital radius on the phase deviation. Section 5 is the discussion and conclusion.

## 2 Specifications of the spacecraft orbit simulator and the space plasma density model

To address the questions raised in the introduction, the NASA General Mission Analysis Tool (GMAT) is used to provide the realistic spacecraft orbit. While several advanced models can give the plasma parameters in the circumstance of the TianQin constellation (Toth, 2005; Shue et al., 1997; Hu. Y. Q. et al., 2005; Xu & Liu, 2014), the Space Weather Modeling Framework (SWMF) model (Toth, 2005) is used to generate the real-time plasma density maps in this work due to its open access policy. This section describes these two tools.

### 2.1 The spacecraft orbit simulator

GMAT is an open access software for modeling and optimizing the spacecraft trajectories. Taking into account the gravity of the celestial bodies among the solar system, Ye (2019) and Tan (2020) recently optimized the orbit of TianQin constellation with GMAT. They have found several stable orbits with the radius of $10^5$ km. These orbital data are used in this paper.

As an example, Figure 2 (a) shows the spacecraft orbit on the whole day of Aug-10-2034. This date is within the detection period, *i.e.* the period of AB. The orbital plane is almost perpendicular to the Sun-Earth line on this day. The shape of the spacecraft is zoomed in for a better view. The velocity of the spacecraft is about 2 km/s at the radius of $10^5$ km. The red curves represent the orbits started from 00:00 UT and terminated at 24:00 UT. SC1, SC2 and SC3 indicate the final positions of the three spacecraft at 24:00 UT. The original orbit adopts the EarthMJ2000Ec coordinate system (GMAT user guide, 2012), where the *x* axis points along the line formed by the intersection of the Earth's mean equator and the mean ecliptic plane at the J2000 epoch, the *z* axis is normal to the mean equatorial plane at the J2000 Epoch. Figure 2 (b) gives another constellation view, indicating the orbit plane on the day of Oct-25-2034, when it is within the non-detection period, *i.e.* the period of BC in Figure 1. The orbital plane is nearly parallel with the Sun-Earth line.

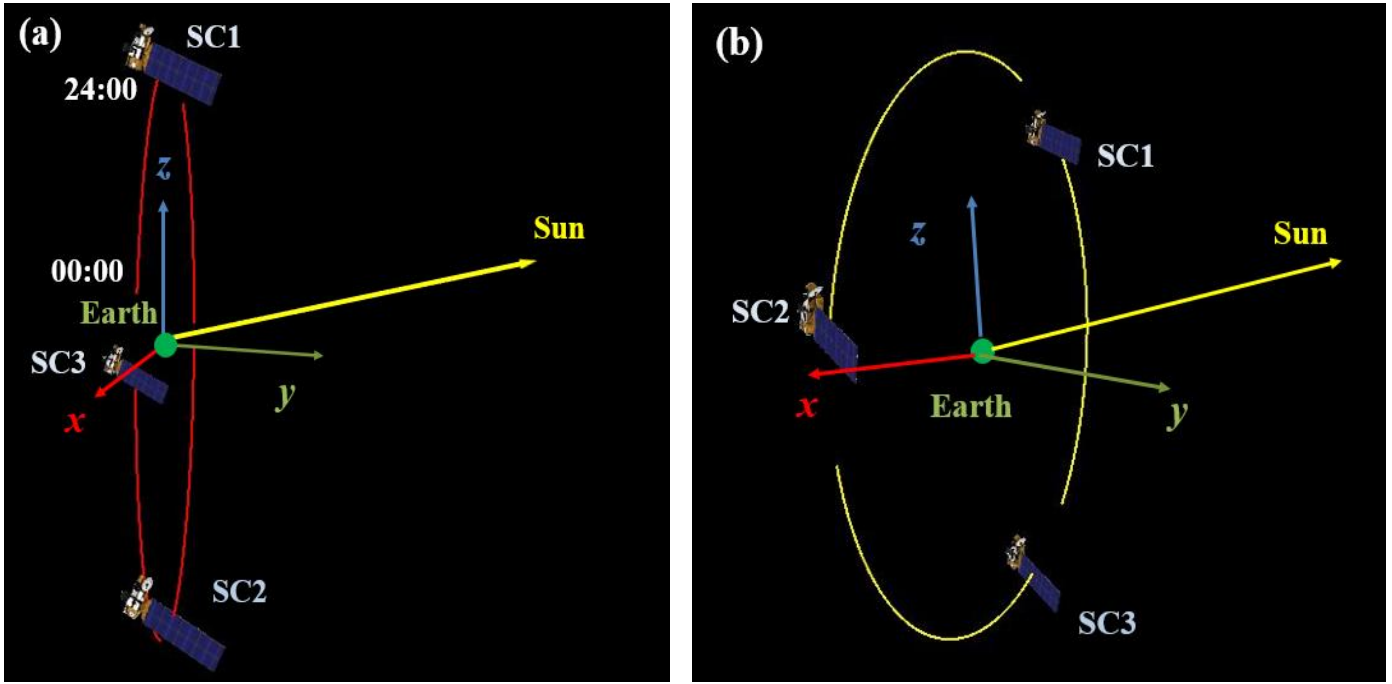

Figure 2. (a) Orbit view of the TianQin constellation on Aug-10-2034, generated by GMAT. Red curves indicate the spacecraft trajectories during the whole selected date. The Earth, the direction to the Sun are shown. (b) Orbit view of the TianQin constellation on Oct-25-2034.

## 2.2 The space plasma density model

The SWMF model (Toth, 2005) was developed by the University of Michigan and offered to the public through the NASA Community Coordinated Modeling Center (CCMC). While modeling the global magnetosphere, it solves the MHD equations with realistic input data from *in-situ* measurements of the solar wind by the Advanced Composition Explorer (ACE) spacecraft (Stone et al., 1998), the Deep Space Climate Observatory (DSCOVR) (Burt & Smith, 2012) and the Wind spacecraft (Ogilvie et al., 1995).

SWMF can generate real time plasma density as well as other common plasma parameters, e.g. magnetic field, current density and temperature. Output data are specified in the Geocentric Solar Magnetospheric (GSM) coordinate (Laundal & Richmond, 2017), in which the origin is the center of the Earth. *x* axis points from the center of the Earth to the Sun and *z* axis is the projection of the Earth's magnetic dipole (positive north) on the plane perpendicular to *x* axis. The global magnetosphere simulation in SWMF covers the region from -250$R_E$ to 33$R_E$ in *x* direction, and from -48$R_E$ to 48$R_E$ in other two directions. Constrained by the input density data from measurement, the temporal resolution is 1min in the finest, whereas the finest space resolution is about ¼ $R_E$ within 6 $R_E$ to the center of the Earth and ½ $R_E$ at the orbital radius of 15.6$R_E$. The real-time solar wind data is projected from ACE spacecraft to the inflow boundary of the simulation region at 33$R_E$.

Since the plasma density simulation relies on the realistic solar wind data from *in-situ* measurement as input, we cannot predict the plasma distribution for TianQin orbit in 2034 right now. Instead, we analyzed the previous real space weather events, as indications for the space weather in 2034. As an example of the magnetospheric plasma density distribution, Figure 3-Figure 4 show the plasma density at 00:00 UT on Oct-24-2016. The red and yellow dashed circles in Figure 3 correspond to two types of the spacecraft orbit alignments selected from the TianQin mission, the detection alignment and non-detection alignment. In reality, the constellation is not necessary inside one plane under the GSM coordinate, due to the tilt angle between the axis of magnetic pole and the axis of rotation of the Earth. This is ignored in the schematic view of the satellite orbits in Figure 3. As seen from Figure 3, the plasma environment is clearly different between orbit 1 and orbit 2.

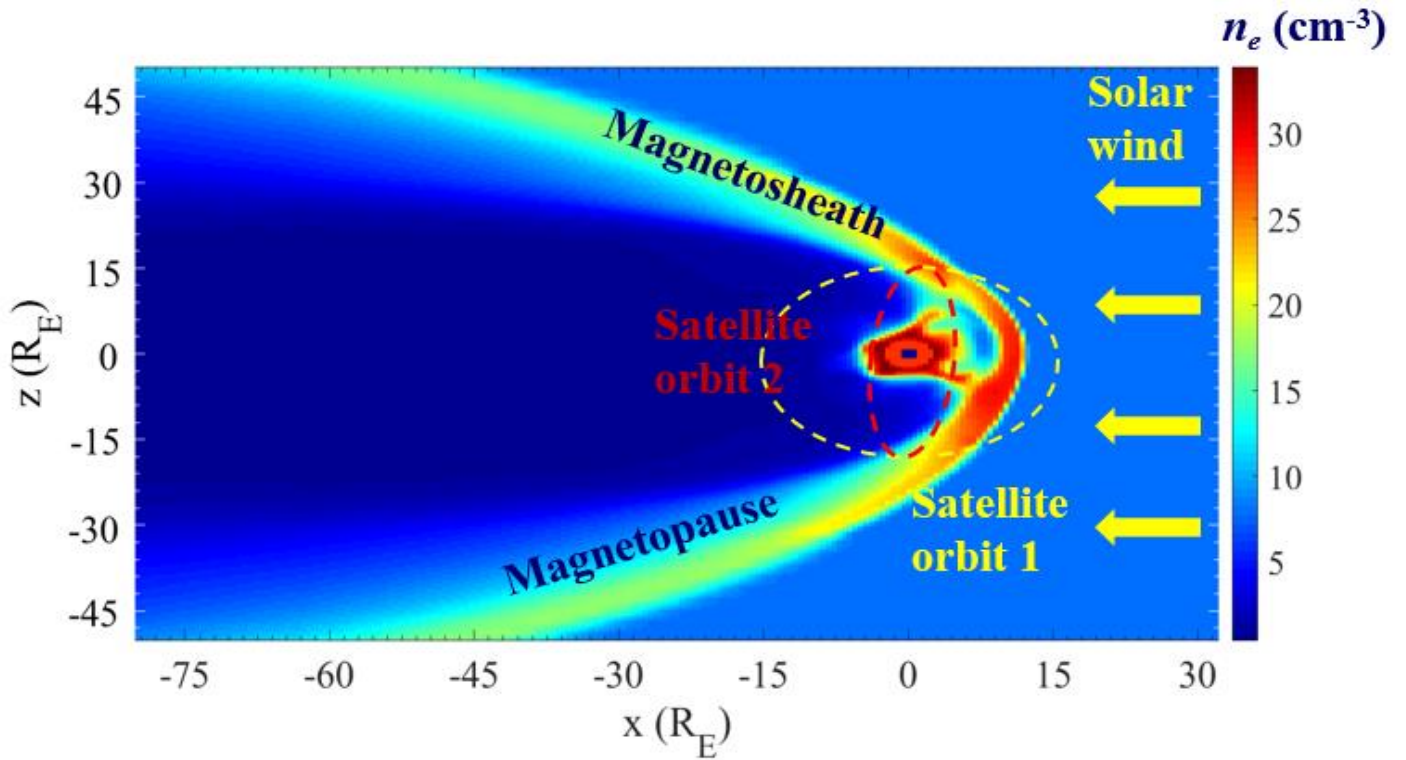

Figure 3. View of the plasma density map in x-z plane, y=0. Data picked at 00:00UT of Oct-24-2016. Two spacecraft orbit alignments are shown in dashed circles. Note the GSM coordinate is used hereafter, which is different from Figure 2.

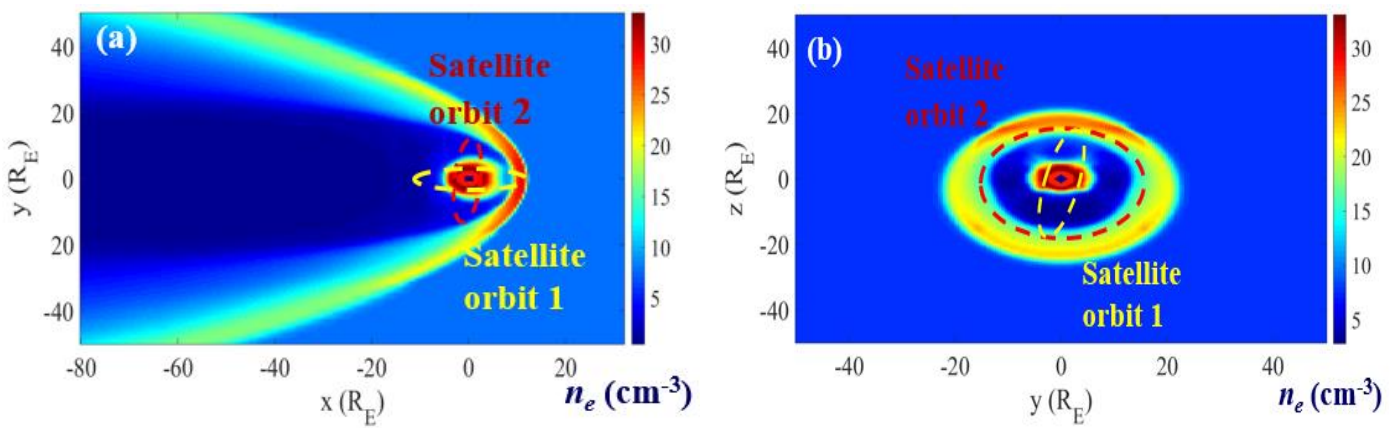

Figure 4. Plasma density model in other planes than x-z at 00:00UT on Oct-24-2016. (a): plasma density profile in x-y plane, z=0; (b): plasma density in y-z plane, x=0. The dashed circles represent two orbit alignments.

# 3 Analysis of the laser's phase deviation caused by real space weather events at TianQin orbit

Magnetic storm is inevitable both in the detection period and the non-detection period of the TianQin constellation. Thus it is necessary to analyze the effects of this space weather events on the laser ranging experiment. For this purpose, two real space weather events are chosen as examples, one minor magnetic storm (6_≥$Kp$ >5_) occurred on Aug-10-2000 and the other moderate magnetic storm (7_≥$Kp$ >6_) on Oct-25-2016. In each event, a specific time period which totally covers the corresponding magnetic storm is selected. Then the SWMF model and the GMAT orbit simulator are used to generate the time-varying plasma density maps and spacecraft positions for each minute in the selected time periods. A time-varying 3D plasma density map is constructed from SWMF data. The density model covers from -20 $R_E$ to 20 $R_E$ in each of the three directions, with a temporal resolution of 1min. The coordinate system of the orbit data is converted from the EarthMJ2000Ec to the GSM, in order to be consistent with the SWMF model. With the three spacecraft positions known, one can fix the propagation path of the three laser arms in each minute. The plasma density map is then interpolated into each laser arm, with 200 grid points in each arm. Finally, the phase deviation can be calculated by Eq. (2) for each minute in both the time domain and the frequency domain. With a standalone laptop computer, a single run for the phase deviation of one minute takes about 3 mins.

## 3.1 Example of the phase deviations in the detection period

Following the same strategy as in our previous work (Lu et al., 2020), we picked the plasma densities along three laser arms from 4am to 10am on Aug-10-2000. A minor magnetic storm with $Kp$ =5 started at 05:20 and persisted until 09:10. The phase deviations of three arms are then calculated by Eq. (2). Figure 5 shows that the value of the phase deviation is in the order of $10^{-6}$ rad at the magnetic quiet time at 04:00-05:20 UT, but suffers a considerable increase during the magnetic storm in-between 05:20 and 09:10 UT. To give a direct comparison to the error budget, the Amplitude Spectral Density (ASD) of this phase deviation is further computed in Figure 6. The black dashed line in Figure 6 marks the error budget of the phase ASD as described by Eq. (1) in the introduction. It reveals that the phase ASD curve is lower than the black dashed line at the milli-hertz frequency band. Higher frequency part in Figure 6 is terminated due to the limited time resolution (1 min) of the density data.

Figure 5-Figure 6 have shown the simulation of the plasma induced phase deviation under a minor magnetic storm. It is important to know how the phase ASD curve in Figure 6 changes under magnetic quiet time or in an even stronger magnetic storm. A moderate magnetic storm is chosen to see the behavior of the phase ASD. This real event occurred during 08:00-11:00 UT on Oct-25-2016. In order to have a direct comparison with Figure 6, the orbit data of Aug-10-2034 is still used, although strictly speaking, the GW detection should be suspended during the time period of this magnetic storm. We picked the plasma densities from 04:00 UT to 11:30 UT on Oct-25-2016 in the SWMF model to cover this magnetic storm and meanwhile to preserve a sufficient amount of data points. The ASD result is then shown in Figure 7. Compared to Figure 6, while still remaining below the error budget, the spectral density curves move upward only slightly at the milli-hertz frequency band. In particular, the ratio of the magnitude of the three-arm averaged phase ASD over error budget at 6mHz raises from 0.132 to 0.2344. However, the ASD curves rise sufficiently around $10^{-4}$ Hz. Since the typical duration of the magnetic storm is several hours, it is reasonable that the increment of the density during the magnetic storm mainly contributes to the sub-mHz frequency band, while having less influence on the mHz frequency band. Following this way of thought, it can be anticipated that the ASD curves would lift a step further than Figure 7 during a major magnetic storm, i.e. staying very close to the error budget.

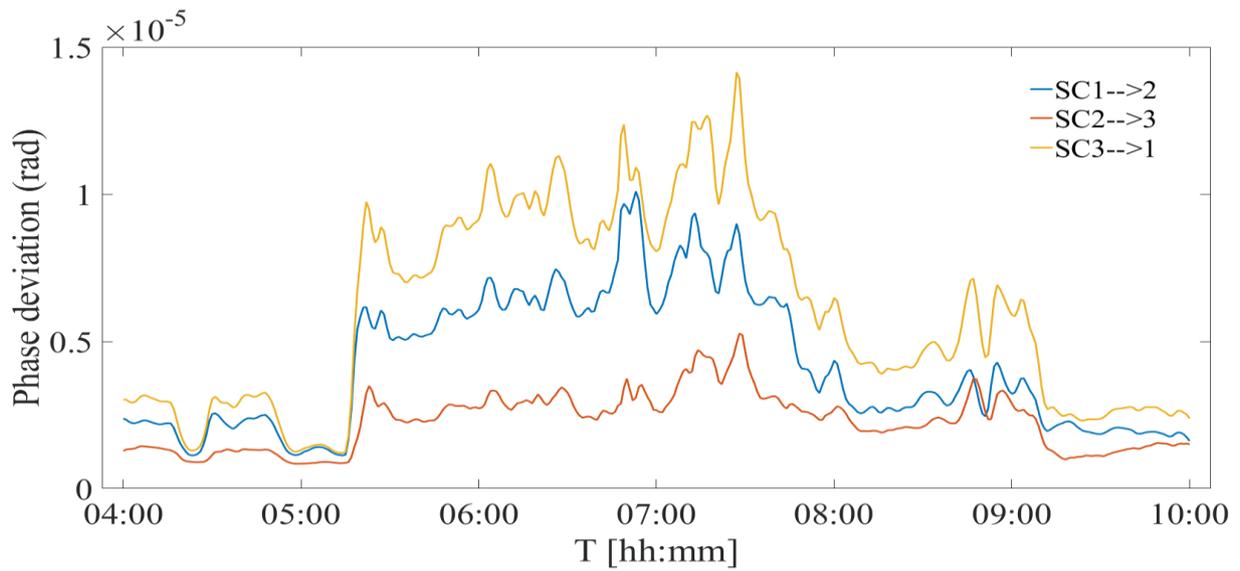

Figure 5. Phase deviations along three arms VS time in the detection period, making use of the plasma density on Aug-10-2000 and orbital data on Aug-10-2034.

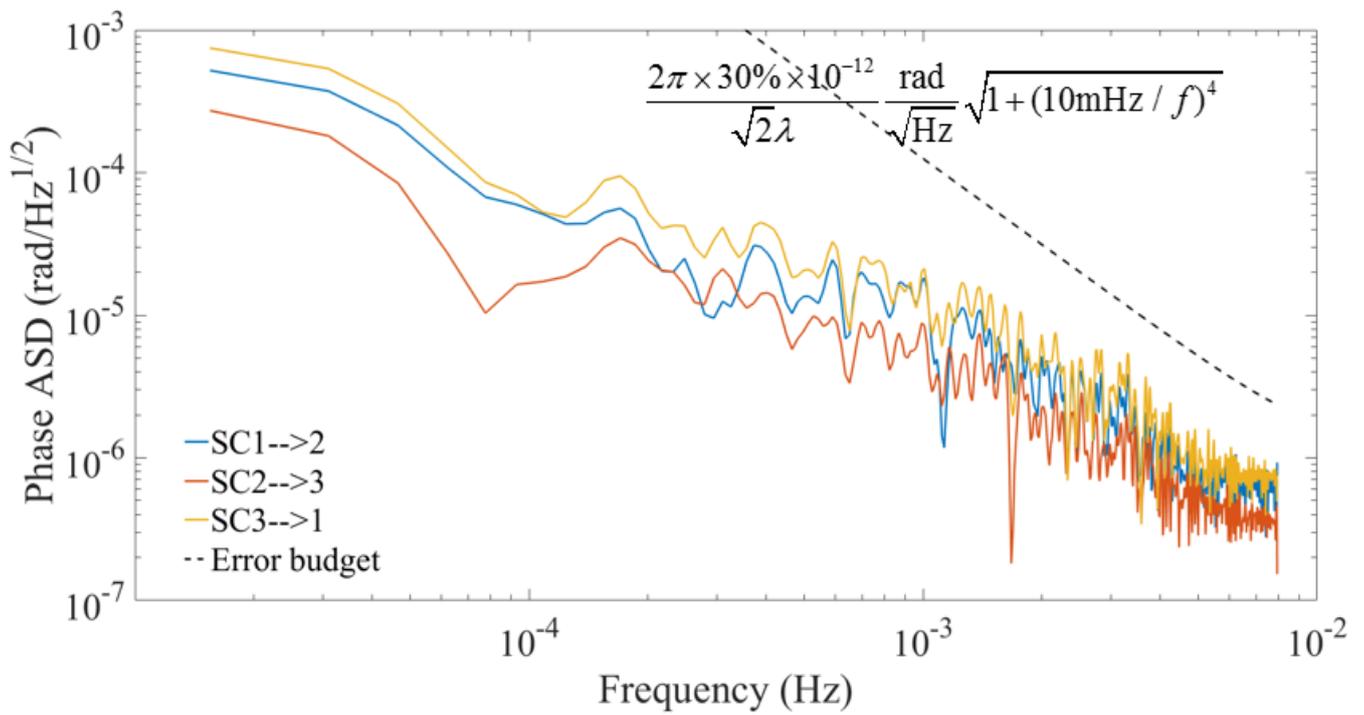

Figure 6. Power spectral analysis of phase deviations calculated in Figure 5. The black dashed curve shows the error budget.

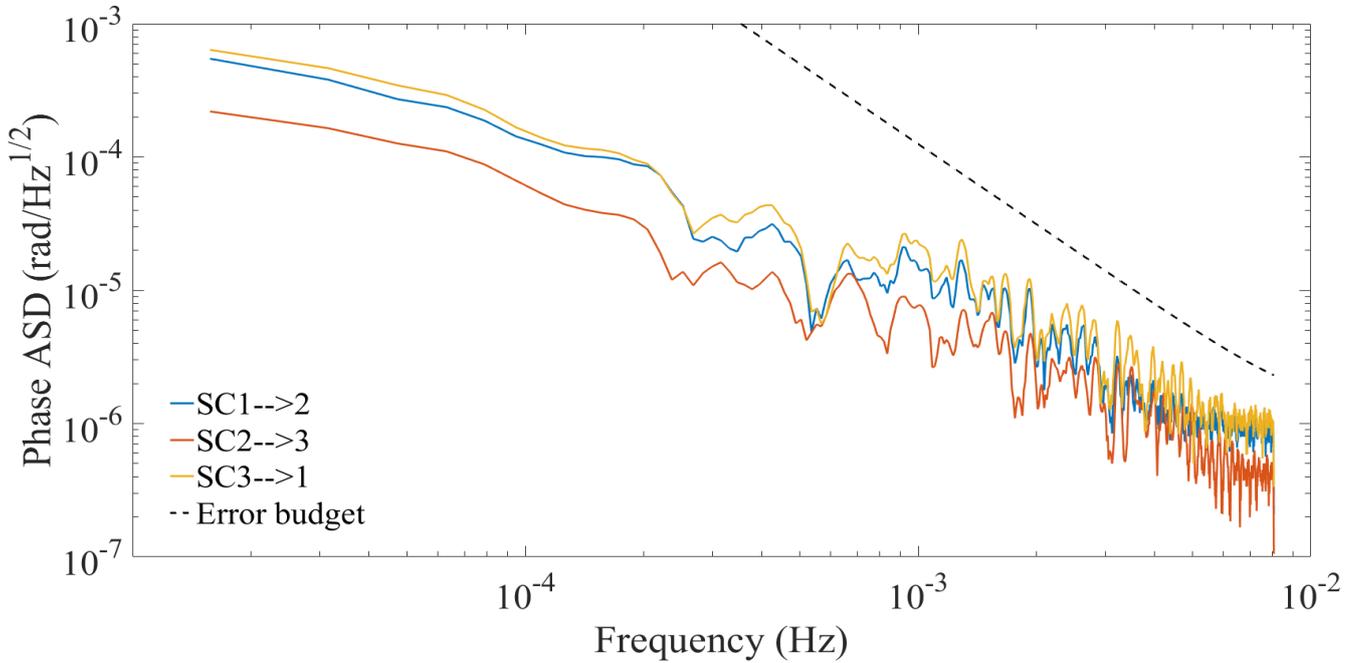

Figure 7. Power spectral analysis of the phase deviations along three arms, using the plasma density on Oct-25-2016 and orbit data on Aug-10-2034. The error budget curve is the same as in Figure 6.

Based on the results in Figure 6 and Figure 7, one may speculate that the spectral curve could be slightly move downward under a magnetic quiet time. Further simulation taking the time period of 00:00-04:00 UT on Aug-10-2000, which has a $Kp<3$ confirms this speculation, see Figure 8. Figure 8 also shows the discrepancy of the phase ASD among three arms is decreasing with mild space weather. This space weather condition is favored by the GW detection since it will not contribute to the converse variation of the phase deviations along two laser arms as the GW could have done.

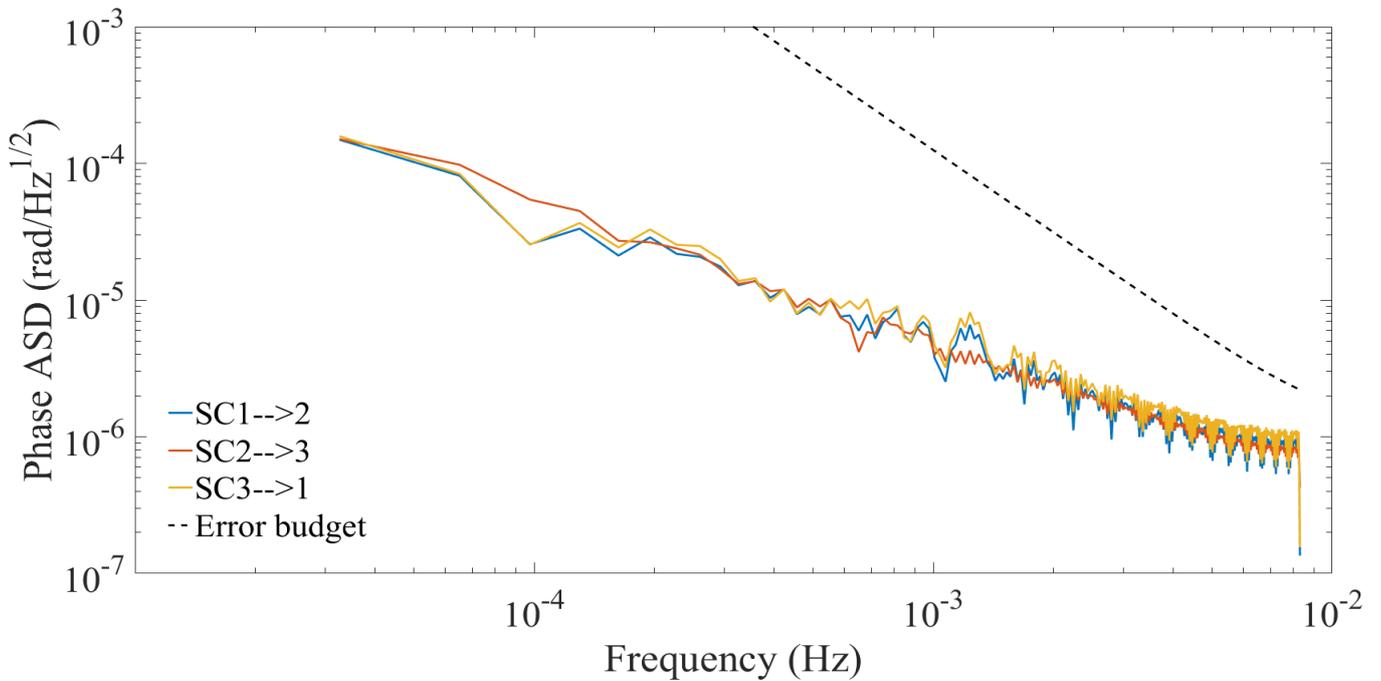

Figure 8. Phase ASD during magnetic quiet time, plasma density data picked in the time interval of 00:00-04:00 UT on Aug-10-2000, orbit data on Aug-10-2034. The error budget curve is the same as in Figure 6 and Figure 7.

One may argue that the uncertainty of the plasma density from the SWMF model could result in the errors of the obtained phase ASD. As an example, error of the phase ASD curve shown in Figure 6 is estimated afterwards. A comparison of the SWMF predicted density data ($n_e^{pre}$) with measured values $n_e^{mes}$ has been done *apriori* by Welling & Ridley (2010). The normalized Root-Mean-Square Error (RMSE) of the plasma density is obtained,

$$nRMSE = \sqrt{\frac{\sum_{i=1}^{N}(n_{e,i}^{pre} - n_{e,i}^{mes})^2}{\sum_{i=1}^{N} n_{e,i}^{mes\ 2}}} \approx 0.961 \qquad (3)$$

From Eq. (2), the phase deviation is linear proportional to the plasma density. Therefore, a crude estimation of the uncertainty of the obtained phase deviation could be,

$$RMSE_{\Delta\phi} \approx \frac{\pi L}{\lambda} \frac{q^2}{\varepsilon_0 m_e \omega_0^2} RMSE_{n_e} \approx \frac{\pi L}{\lambda} \frac{q^2}{\varepsilon_0 m_e \omega_0^2} \bar{n}_{e,i}^{mes} \times nRMSE \approx 4.3 \times 10^{-13}\ \bar{n}_{e,i}^{mes}\ rad \qquad (4)$$

Where $\bar{n}_{e,i}^{mes}$ is the averaged plasma density from measurement. Depending on the space weather, it may vary from $10^6$ m$^{-3}$ (during magnetic quiet time, i.e. 04:00-05:20UT and 09:10-10:00UT) to $10^7$ m$^{-3}$ (during a magnetic storm, i.e. 05:20-09:10UT). Substitute Eq. (4) into the ASD calculation, the upper limit of the ASD curves due to the density uncertainty can be assessed. Figure 9 shows the upper limit of the yellow phase ASD curve of Figure 6. The result indicates the phase ASD could shift upwards by $10^{-6}$ rad/Hz$^{1/2}$ in the mHz frequency band.

In terms of the moderate or major magnetic storm, the uncertainty of the density data could presumably become even larger, which would lead to the phase ASD eventually exceeding the error budget. It is strongly recommended that the GW detection should be paused during these space weather events.

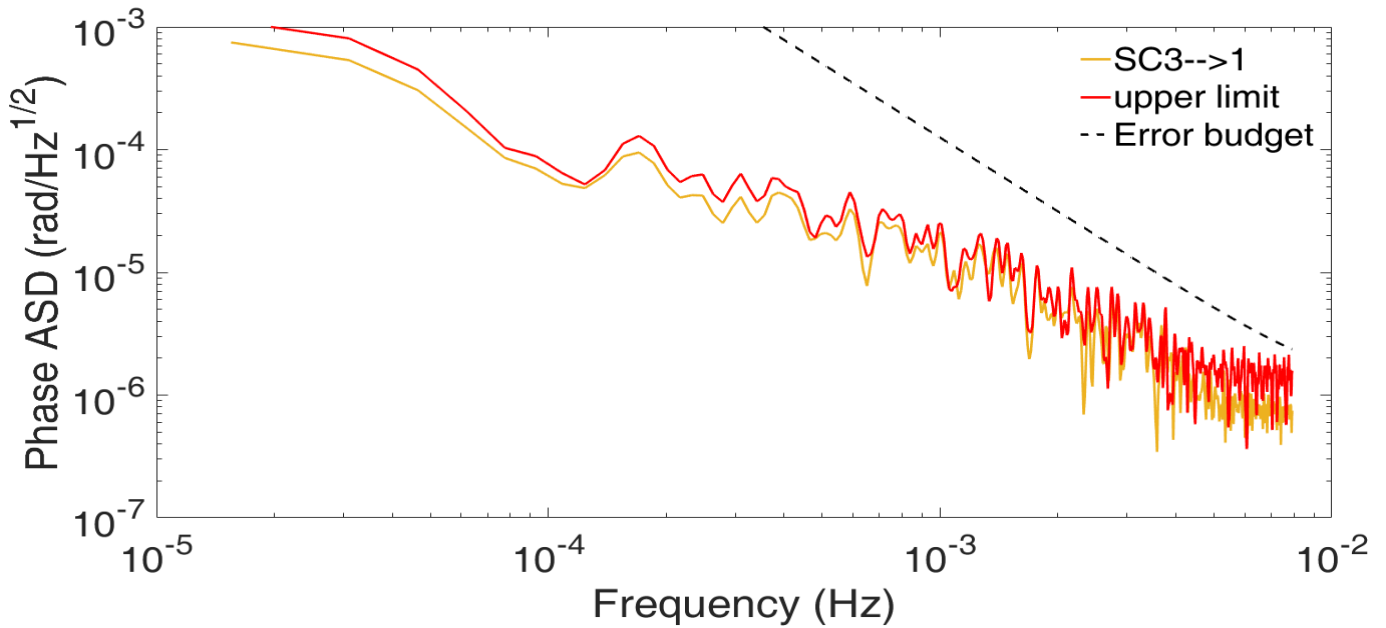

Figure 9. Error analysis for yellow curve of Figure 6. The yellow solid line and black dashed line are the same as in Figure 6. The red solid line indicates the upper limit of the yellow solid line due to the density uncertainty.

## 3.2 Example of the phase deviations in the non-detection period

Furthermore, simulations are done to investigate the plasma induced phase deviations in the non-detection period. This section uses the same plasma density that has been used in Figure 7, *i.e.* densities on Oct-25-2016, in order to have a better comparison. Accordingly, the orbit data on Oct-25-2034 is chosen. Both the plasma density and the orbit data are picked from 01:00 to 11:30 UT at the corresponding dates. The temporal variations of the phase deviation along three arms are shown in Figure 10. The blue and yellow lines behave conversely between 01:00 and 08:00, as one arm comes out of the magnetosheath whereas the other gets inside during this period. The red line is sufficiently lower than the other two as most part of this arm lies in the magnetotail side of the magnetosphere. The new ASD is shown in Figure 11. Probably due to the discrepancy of the plasma environment between the magnetopause side and magnetotail side, the three curves in Figure 11 show larger separation than those in Figure 6-Figure 8. As a result, with this orbit alignment, the plasma induces larger phase difference between different laser arms than in the detection orbit. This result along with the sunlight contamination problem, provide justifications that one should better suspend the GW detection under this orbit alignment.

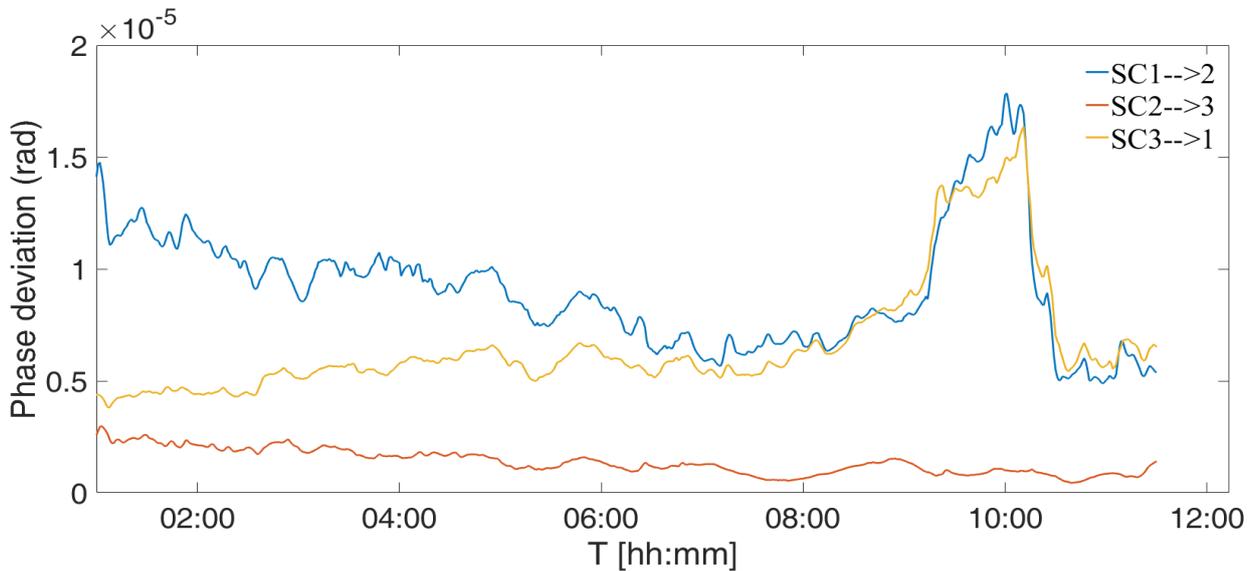

Figure 10. Time variation of phase deviations along three arms in the non-detection period, using the plasma density on Oct-25-2016 and orbit data on Oct-25-2034.

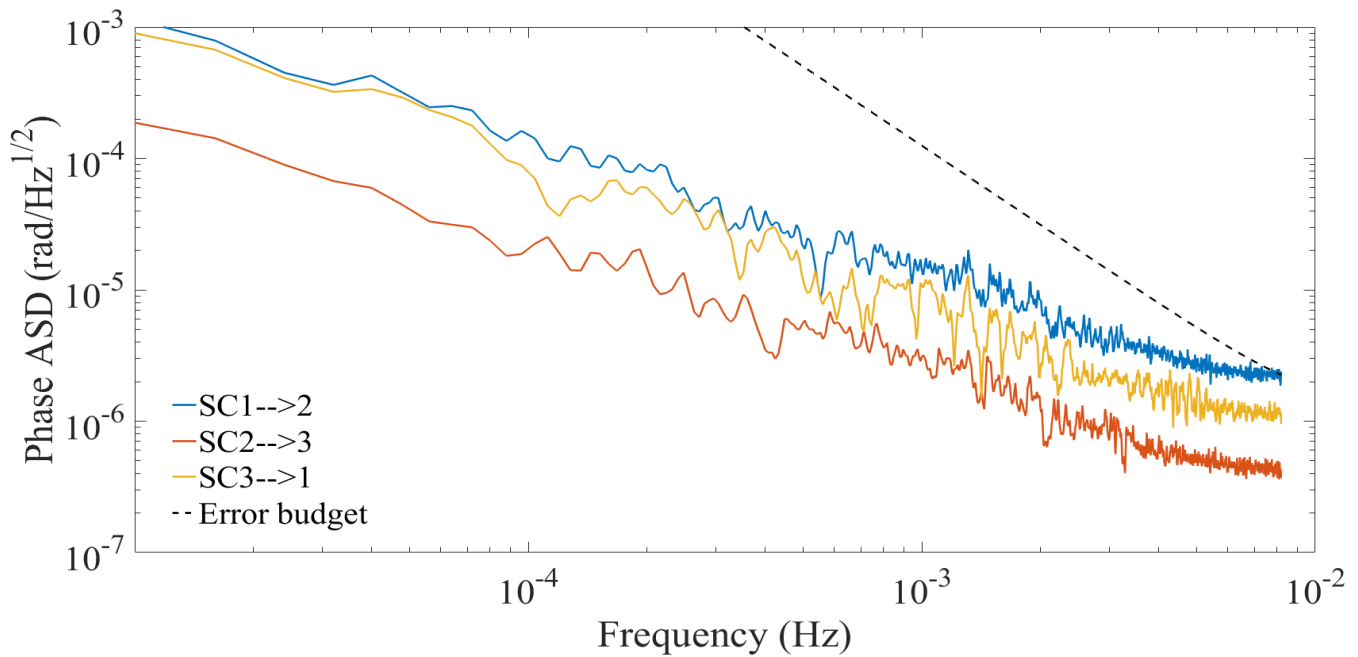

Figure 11. Power spectral analysis of phase deviations in Figure 10. The black dashed line is the same as in Figure 8.

## 4 Examination of plasma induced phase deviations at different orbital radii

This section aims to find the correlation between the phase ASD and the orbital radius. Radii between $7\times10^4$ km ($11R_E$) and $15\times10^4$ km ($23R_E$) are examined in this section for this purpose. Selection of the orbit is constrained by the stabilization of the orbit and the sensitivity of the GW detection. The orbit will become unstable under larger radius (Tan et al., 2020). The GW sensitivity of a Michelson interferometer represents the probability of the successful detection of the GW wave. A 'high' sensitivity is preferred in order to detect as many as possible the GW sources. Lower radius (shorter arm length) will reduce the GW sensitivity in the milli-hertz (Larson et al., 2000) frequency band. Moreover, it is worth mentioning that the error budget for the phase deviation is linearly proportional to the arm length. In case of $15\times10^4$ km ($7\times10^4$ km), the error budget will become 1.5 (0.7) times the criterion of $10^5$ km.

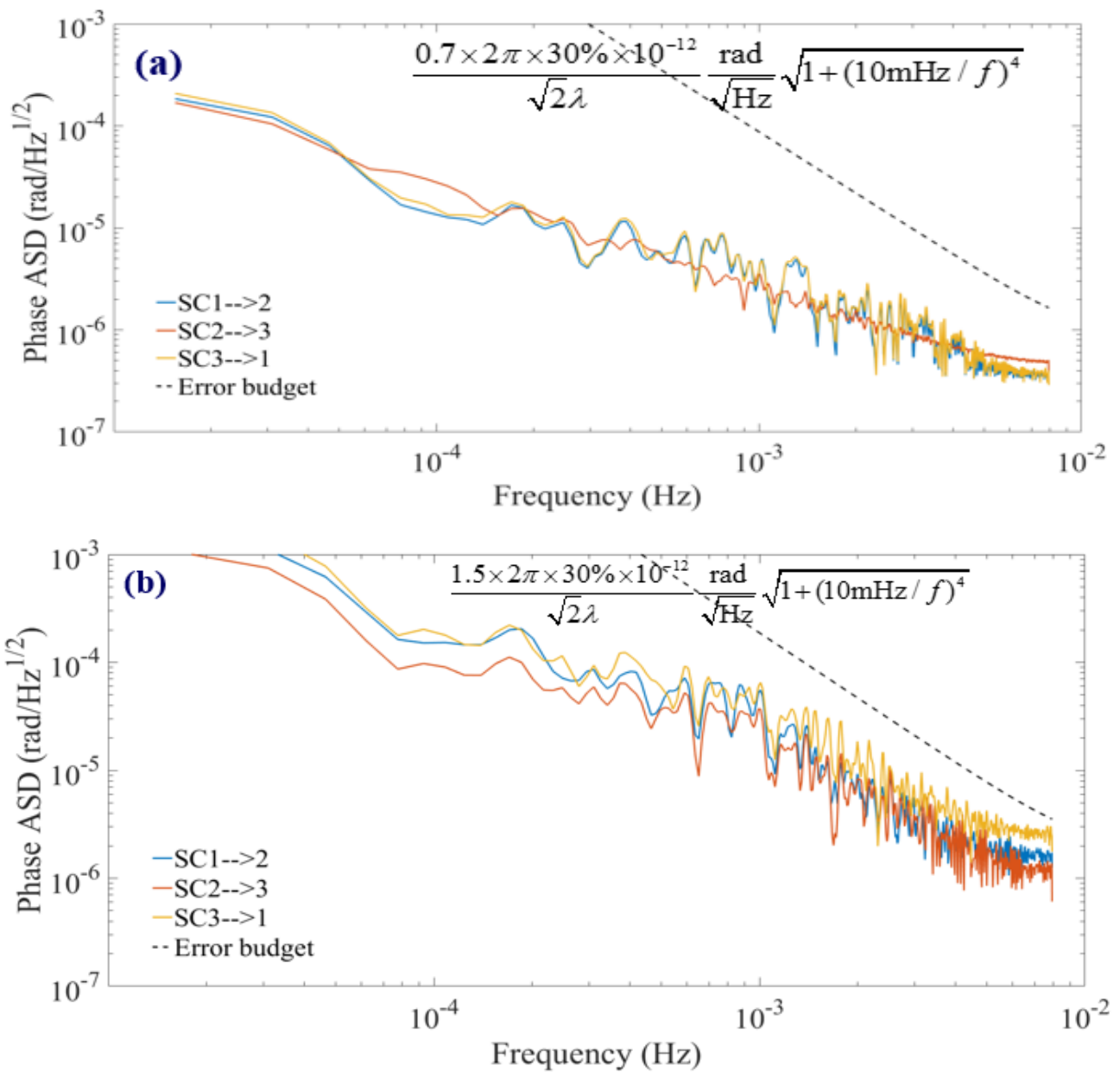

Figure 12. (a) The same data period as Figure 6, but with orbital radius of $7\times10^4$ km; (b) orbital radius of $15\times10^4$ km. The black dashed lines at these two figures show the updated error budget.

Figure 12 draws the phase spectra with two different orbital radii, $7\times10^4$ km and $15\times10^4$ km, respectively. The plasma densities are picked from 04:00 to 10:00 on Aug-10-2000 as in Figure 6. The Results show the phase ASD decreases as the radius decreasing. This is at first sight counter intuitive since the periodic oscillations of the plasma density would be significantly reduced if the spacecraft move out of the magnetosheath, where the plasma density is higher than its vicinity. On the other hand, from Eq. (2), the phase deviation could be indeed decrease with a shorter arm length, if the plasma density does not change significantly with the variation of the altitude of the orbit. The cumulative effects of the long distance seems to be more dominant in determining the phase deviation. As a consequence, Figure 12 indicates that even the error budget is more stringent under lower radius, reduce the radius could help suppress the phase deviation to some extent. In fact, the ratio of the magnitude of the three-arm averaged phase ASD over error budget at 6mHz raises from 0.17 to 0.41 when the orbit radius increases from $7\times10^4$ km to $15\times10^4$ km. Purely from this point of view, reducing the orbital radius is a way to mitigate the phase deviation.

## 5 Discussion and Conclusion

This paper reports the preliminary results regarding to the effects of the plasma density oscillations on the phase deviations of the lasers for the TianQin GW observatory. The calculations are based on the realistic constellation orbit and real-time space plasma density model. According to the current plasma model, the phase deviation due to the plasma density is within the designated error budget. Besides, the discrepancy of the phase deviations among three arms is more significant when the orbital plane is parallel to the Sun-Earth line than the perpendicular alignment. Thus even from this point of view, one would better pause the GW detection during the parallel orbit alignment. Moreover, magnetic storm could also exacerbate the phase discrepancy among three arms, although the most vulnerable frequency band is sub-mHz. This emphasizes the importance of the surveillance and forecasting the space weather and the solar activity during the GWs detection. Simulation also shows that reducing the orbital radius could mitigate the phase deviation to a certain extent.

It is worth mentioning that our results rely on both the *in-situ* measurement and modeling of the plasma parameters. Taking the density uncertainty into consideration, our results show the maximum phase ASD is still below the error budget during the mild space weather condition, whereas a moderate or major magnetic storm could be problematic. More rigorous error analyses and validation of the results are needed, e.g. comparison between different density models. Moreover, current results of the phase ASD neglect the high frequency oscillations of the plasma density due to the limited resolution of the input density data from measurement. This can be improved when a more advanced model and finer *in-situ* measurement are available.

**Acknowledgment**

The authors wish to thanks Prof. Chen Zhou and Prof. Jian-Wei Mei for many useful discussions. The SWMF Model was developed by the Center for Space Environment Modeling at the University of Michigan. Some simulation results in this paper have been provided by the Community Coordinated Modeling Center at Goddard Space Flight Center through their public Runs on Request system (http://ccmc.gsfc.nasa.gov). The GMAT software was accessed from https://software.nasa.gov/software/GSC-17177-1.This work has been supported by the China Postdoctoral Science Foundation under grant agreement 2018M643286, the postdoctoral funding project of the Pearl River Talent Plan, the National natural Science Foundation of China under grant agreements 11805287, 11803008 and 42004156.